\documentclass[11pt,tightenlines,eqsecnum,floats,aps,amsmath,amssymb,nofootinbib,prd,shownopacs,floatfix]{revtex4-2}
\usepackage{comment}

\usepackage{graphicx}

\usepackage{epstopdf}
\usepackage{latexsym}
\usepackage{amssymb}
\usepackage{amsmath}
\usepackage{color}
\usepackage{mathrsfs}
\usepackage{xparse}
\usepackage{float}
\usepackage{mathtools}
\usepackage{multirow}

\usepackage[table,xcdraw]{xcolor}

\usepackage[normalem]{ulem}
\useunder{\uline}{\ul}{}

\usepackage[center]{subfigure}

\begin{document}
\renewcommand\arraystretch{2}
 \newcommand{\bq}{\begin{equation}}
 \newcommand{\eq}{\end{equation}}
 \newcommand{\bqn}{\begin{eqnarray}}
 \newcommand{\eqn}{\end{eqnarray}}
 \newcommand{\nb}{\nonumber}

 \newcommand{\cb}{\color{blue}}

    \newcommand{\cc}{\color{cyan}}
     \newcommand{\lb}{\label}
        \newcommand{\cm}{\color{magenta}}
\newcommand{\rc}{\rho^{\scriptscriptstyle{\mathrm{I}}}_c}
\newcommand{\rd}{\rho^{\scriptscriptstyle{\mathrm{II}}}_c}
\NewDocumentCommand{\evalat}{sO{\big}mm}{%
  \IfBooleanTF{#1}
   {\mleft. #3 \mright|_{#4}}
   {#3#2|_{#4}}
}

\newcommand{\PRL}{Phys. Rev. Lett.}
\newcommand{\PL}{Phys. Lett.}
\newcommand{\PR}{Phys. Rev.}
\newcommand{\CQG}{Class. Quantum Grav.}
\newcommand{\parallelsum}{\mathbin{\!/\mkern-5mu/\!}}

\title{Cyclic universe and uniform rate inflation from polymerized scalar field in loop quantum cosmology}
\author{Bikash Chandra Paul $^{1}$}
\email{bcpaul@nbu.ac.in}
\author{Sahil Saini $^{2}$}
\email{sahilsaiini@gjust.org}
\affiliation{$^{1}$ Department of Physics, University of North Bengal, Siliguri, Darjeeling 734 013, West Bengal, India\\
$^{2}$ Department of Physics, Guru Jambheshwar University of Science $\&$ Technology, Hisar 125001, Haryana, India\\}

\begin{abstract}

We investigate the uniform rate inflationary universe in the framework of loop quantum cosmology (LQC) and find that this seemingly simple inflationary model is interlinked with various concepts such as cyclic evolution, HNI inflation, and polymer-quantized scalar fields when the background spacetime is loop-quantized. The potential for an \textit{exactly} uniform rate inflation in a loop-quantized FRW spacetime turns out to be a polymerized version of the corresponding potential in general relativity, which mimics the potentials for a polymerized scalar field and hybrid natural inflation (HNI). There is also a radical modification of the background spacetime, leading to a cyclic universe with identical epochs separated by quantum bounces that replace the classical singularity. The energy density and Hubble rate are bounded. Polymerization of the scalar field is found to be intimately tied to the polymerization of the background geometry in LQC. Remarkably, a single polymerized scalar field simultaneously governs the cyclic evolution of the universe and drives the inflationary phase. Preliminary predictions for cosmological perturbations are worked out and are found to depend on the value of the field at the end of inflation. The parameter space is explored to compare the results for the spectral index and tensor-to-scalar ratio with observational constraints.
\end{abstract}

\maketitle

\section{Introduction}
\label{intro}

Cosmic inflation is a widely accepted model for the early universe \cite{1,linde,1a}. While originally introduced to address several issues in big bang cosmology, it also provided a causal mechanism for the origin of structure in the universe that is largely consistent with current observations \cite{2,Planck}. Quantum fluctuations during the inflationary expansion seeded the scalar and tensor perturbations which left their imprints on the cosmic microwave background radiation (CMB), which is observed today with high precision, thus providing an effective method to impose constraints on cosmological model parameters \cite{Planck}. Although numerous inflationary models have been developed, the precise onset of inflation remains uncertain. Among them, Linde’s chaotic inflation is among the simplest models which remains a key framework, requiring an initial scalar field value $\phi > 3 M_{\mathrm Pl}$ \cite{linde,lin}, and is valid even in anisotropic \cite{bc} and braneworld contexts \cite{bc1}. Another major problem is that inflationary spacetimes are past-incomplete \cite{BGV}. In other words, inflation does not resolve the big bang singularity. The present work considers uniform rate inflation, which is based on a uniformly rolling inflaton field, in loop quantum cosmology. Uniform rate inflation offers an exactly solvable and observationally consistent framework, previously studied in general relativity \cite{5}, braneworld cosmology \cite{6} and $f(T,\mathcal{T})$- gravity \cite{11BCP}. The present work extends this analysis to loop quntum cosmology which provides a singularity free background spacetime.

Loop quantum cosmology is an application of the canonical quantization methods of loop quantum gravity to cosmological spacetimes \cite{lqcreview1,lqcreview2}. Loop quantization reveals a microscopic and discrete underlying geometry for the spacetime whose dynamics are described by quantum difference equations. When applied to the FRW spacetime, it leads to a resolution of the big-bang singularity without requiring exotic matter fields or the violation of energy conditions, while also leading to a universal and Planckian upper bound on the energy density of the universe called the critical energy density, $\rho_c \approx 0.41 \rho_{\mathrm Pl}$ \cite{aps}. The cosmological singularity is replaced by a quantum bounce in loop quantum cosmology (LQC) which occurs at a finite volume when the universe reaches the critical energy density. Interestingly, an effective continuum description can be obtained for LQC spacetimes which incorporates the most important underlying quantum corrections to the dynamics and provides a convenient arena for phenomenological studies using differential equations. These quantum corrections lead to modified Friedmann and Raychaudhuri equations which describe the dynamics of the effective spacetime. This effective dynamics has been tested for various isotropic as well as anisotropic spacetimes to faithfully approximate the underlying quantum dynamics when starting from macroscopic initial conditions for states sharply peaked on classical trajectories \cite{Diener:2013uka}. Using the effective dynamics, it has been shown that the effective spacetimes of various isotropic and anisotropic models in loop quantum cosmology are free from all strong curvature singularities and are geodesically complete irrespective of the choice of the matter content \cite{Singh:2009mz,Saini_thesis,Saini:2016vgo}. Furthermore, the effective equations for the background spacetime have been found to approximate the underlying discrete geometry to a very high accuracy even in presence of cosmological perturbations in a loop quantized isotropic spacetime with a massive inflaton field \cite{AshtekarNeilson}. Nevertheless, the corrections in the Mukhanov-Sasaki equation for perturbations are non-negligible. Thus, effective dynamics has been extensively used to model the LQC background in studies of inflationary models embedded in an LQC background. The quantum corrections quickly die down when the energy density falls below Planckian values after the bounce. Thus, inflationary models of one's choice can be readily embedded in an LQC background spacetime with the bounce leaving only tiny effects in the post-bounce inflationary phase. Nevertheless, the combined effect of modified Friedmann equation describing the background as well as the corrections in the Mukhanov-Sasaki equation leave a distinct imprint in the CMB, which has been extensively studied in the literature. Loop quantized spacetimes have been extensively used in the past decade to probe early universe physics in combination with both inflationary models as well as alternatives to inflation, leading to results consistent with current observations yet containing interesting predictions for future observations \cite{LQC_CMB}. In this work, we use the effective dynamics of LQC to study the early universe with a uniform rate inflaton field. 

Using the modified Friedmann equation and the Klein-Gordon equation for the scalar field in loop quantum cosmology, we obtain the potential which yields a uniform rate of evolution for the inflaton field. The quantum geometry effects incorporated through the modification in the Friedmann equation lead to imporatnt consequences. In particular, the potential for an \textit{exactly} uniform rate inflation in isotropic loop quantum cosmology turns out to closely resemble the potentials in case of configuration variable based polymer quantization of the scalar field \cite{polymerfield} and the hybrid natural inflation (HNI) models \cite{HNI1,HNI2}, indicating that both HNI potential and polymer quantized potential for scalar field in a LQC background spacetime admit a solution where the inflaton field evolves at a uniform rate. Moreover, the energy density is oscillatory and bounded as is generically true in LQC. The Hubble rate is also bounded as is generically true in LQC. In addition, the Hubble rate is also oscillatory and we find that the phase space is compactified as found earlier in the case of polymerized scalar field potential in \cite{polymerfield}. However, unlike the work \cite{polymerfield} which considers polymerized potential in classical gravity where the singularity is not resolved, we find in our case that the singularity is resolved due to quantum gravitational effects coming from LQC. Interestingly, we find that uniform rate inflation in loop quantum cosmology leads to a cyclic universe with identical cycles of expansion and contraction separated by the quantum bounces which occur at the critical density. Unlike the majority of studies involving inflation in an LQC spacetime, the bounce in this case is dominated by the potential energy. 

Interestingly, we find that the polymerization effects in the scalar field in this case are dynamically intimately tied to the quantum geometry effects in the background geometry. The dynamical system is not simply a generic polymerized scalar field in an effective LQC spacetime. Rather, the condition of a uniformly evolving scalar field selects a specific polymerization of the scalar field, in which the evolution of the effects of the polymerization very closely follow the quantum geometric effects in the background spacetime which are manifest in the polymerization of the connection. This causes the global evolution to be cyclic, and the quantum effects in both matter and geometric sector rise and fall in tandem as the universe passes through cycles of bounce, expansion, recollapse and contraction. In our study, this single polymerized scalar field which governs the cyclic nature of the global evolution, is also responsible for driving the post-bounce inflationary phase. Since the framework for perturbations of a polymer quantized scalar field on a discrete underlying LQC geometry is not yet fully developed, we calculate preliminary estimates of CMB parameters in this manuscript by including only the effects of the modifications in the background dynamics due to LQC. For these estimates, we use the $\delta N$ formalism. Unlike the case of uniform rate inflation in braneworld cosmology \cite{6} where the brane tension $\Lambda$ can be adjusted to fit the observations, the corresponding parameter in equations in case of LQC, namely the critical energy density in LQC is already fixed in the literature using black hole entropy calculations, i.e. $\rho_c \approx 0.41 \rho_{\mathrm Pl}$. Thus, the amplitude of the observed CMB scalar power spectrum leads to fixing of the parameter $\lambda$, which represents the uniform rate of evolution of the inflation field (in the context of polymer quantized scalar field potential, it would relate to the lattice spacing in polymer quantization \cite{polymerfield}). The values of cosmological parameters such as the spectral index and the tensor-to-scalar ratio depend on the field value at the end of inflation. We explore the parameter space and find that a viable model of uniform rate inflation in loop quantum cosmology can be obtained which satisfies the observation constraints. Due to the fact that quantum modifications quickly become negligible in the inflationary phase after the bounce as soon as the energy density falls below Planckian levels, we find that the predictions for CMB parameters is similar to the predictions from uniform rate inflation in GR. 
Curiously, since the quantum corrections become negligible in the inflationary phase, the potential needed for uniform rate inflation in GR can also be directly used in an LQC background to obtain a nearly uniform rate inflation and similar predictions for CMB parameters. This is the reason most studies on inflation in LQC have embedded the same inflationary potentials in LQC spacetimes as used in GR leading to similar, though not exactly same dynamics for the inflaton. However, as noted above, demanding the inflaton to evolve with an \textit{exactly} uniform rate in LQC spacetime leads to a radical modification in the potential - the potential becomes polymerized version of its classical counterpart. This is the first time that polymerization of matter quantities has emerged in LQC from purely phenomenological considerations. In addition, we find that embedding uniform rate inflation in LQC radically alters the global evolution which becomes cyclic, while having only a tiny effect on the inflationary parameters. Quite surprisingly, we find that various concepts -- namely, HNI inflation, polymerized quantized scalar field, uniform rate inflation and cyclic evolution, which are unrelated in the context of classical gravity, turn out to be closely related when the background spacetime is loop quantized.

The remainder of this manuscript is organized as follows. In section II, we provide a brief review of uniform rate inflation in general relativity. In section III, we consider uniform rate inflation in loop quantum cosmology and obtain the evolution of the universe, as well as various parameters such as the spectral index of the power spectrum and the tensor-to-scalar ratio which are compared with observational constraints. We summarise our results in section IV.

\section{Uniform Rate Inflation in General Relativity}

In this section, we review the uniform rate inflationary model in general relativity. Einstein's field equations are given by
\begin{equation}
\label{1}
    R_{\mu\nu}-\frac{1}{2}  g_{\mu\nu} R = 8 \pi G \; T_{\mu\nu},
\end{equation}
where $G$ is the gravitational constant and $ T_{\mu\nu}$ the energy-momentum tensor, which for an isotropic fluid takes a diagonal form $(\rho, -p,-p,-p)$. In the following, we set the Planck mass to unity, i.e. $M_{\mathrm Pl}=1/\sqrt{8\pi G}=1$. For a homogeneous scalar field $\phi$, the energy density and pressure are given by:
\begin{equation}
\label{2}
    \rho = \frac{1}{2} \dot{\phi}^2 + V(\phi), \; \; \; p = \frac{1}{2} \dot{\phi}^2 - V(\phi).
\end{equation}
We consider a flat and isotropic spacetime described by the FRW metric
\begin{equation}
\label{3}
    ds^2= -dt^2+ a^2(t) \left[dr^2 +r^2(d\theta^2+sin^2\theta d\phi^2)\right].
\end{equation}
For the FRW metric, the field equations simplify to the following Friedmann and Raychaudhuri equations respectively
\begin{equation}
\label{4}
    3\frac{\dot{a}^2}{a^2}= \left( \frac{1}{2} \dot{\phi}^2 + V(\phi)\right),
\end{equation}
\begin{equation}
\label{5}
    2 \frac{\ddot{a}}{a} + \frac{\dot{a}^2}{a^2}= - \left( \frac{1}{2} \dot{\phi}^2 - V(\phi)\right),
\end{equation}
while the conservation equation is given by
\begin{equation}
\label{6}
\ddot{\phi} + 3 \frac{\dot{a}}{a} \dot{\phi} +\frac{dV}{d\phi}=0.
\end{equation}
For a flat universe, a uniform rate of inflation can be obtained from the potential \cite{5} which is given by
\begin{equation}
\label{7}
V(\phi) = \frac{3 \lambda^2 \phi^2}{4} - \frac{\lambda^2}{2}.
\end{equation}
The potential is determined by a quadratic scalar field with a negative cosmological constant. An exact solution to equation \eqref{6} is given by
\begin{equation}
\label{8}
\dot{\phi} = - \lambda,
\end{equation}
corresponding to the inflaton field rolling at a constant speed. The above equation is readily integrated to yield 
\begin{equation}
\label{9}
\phi = - \lambda \, t +\phi_o,
\end{equation}
where $\phi_o$ is an integration constant. 
From equation (\ref{4}), the Hubble rate $H=\dot a /a$ is given as
\begin{equation}
\label{14}
3H^2 = \frac{1}{2} \dot{\phi}^2 + V(\phi) = \frac{3\lambda^2 \phi^2}{4},
\end{equation}
which yields the following expression for the Hubble rate
\begin{equation}
    H= -\frac{\lambda^2}{2}t + \frac{\lambda}{2} \phi_o.
\end{equation}
We define the scale factor to be $a \equiv a_0 e^\alpha$ such that $H= \dot \alpha$. Integrating the above equation yields 
\begin{equation}\label{alpha_GR}
    \alpha=-\frac{\lambda^2 t^2}{4} + \frac{\lambda \phi_o t}{2} + \alpha_0
\end{equation}
Setting $\alpha(0)=\alpha_0=0$, the scale factor is given by
\begin{equation}
    a= a_0 e^{-\frac{\lambda^2 t^2}{4} + \frac{\lambda \phi_o t}{2}}.
\end{equation}
This model is called the uniform rate inflation, where the slow-roll condition is automatically satisfied during inflation. The uniform rate inflation is an attractor solution for the potential \eqref{7}, and this simple model can be made to satisfy the latest observational constraints \cite{5}. It was later implemented in a braneworld scenario \cite{6} and $f(T,\mathcal{T})$-gravity \cite{11BCP}. In the next section, we use a similar methodology to construct a uniform rate inflationary model in loop quantum cosmology.

\section{Uniform rate inflation in loop quantum cosmology}
In this section, we work with the effective dynamics of loop-quantized FRW spacetime with a minimally coupled massive scalar field in the so-called $\bar\mu$-scheme \cite{lqcreview1}. Although the underlying quantum geometry in LQC is discrete, the dynamics of which are described by quantum difference equations, it is possible to obtain an effective continuum description for phenomenological explorations that incorporate the underlying quantum corrections. Effective dynamics has been shown to faithfully approximate the underlying discrete evolution in various isotropic and anisotropic loop-quantized spacetimes, including in the bounce regime \cite{Diener:2013uka}. Even in the presence of cosmological perturbations in an isotropic loop quantized spacetime coupled with a massive inflaton field, the corrections in the effective metric for the background spacetime are found to be extremely small (though the corrections in the Mukhanov-Sasaki equation for perturbations are non-negligible, which lead to distinct signatures in the CMB) \cite{AshtekarNeilson}. Thus, the effective equations for the isotropic loop quantized spacetime used in this manuscript are standard in the literature. 
Moreover in LQC, the quantum corrections are only non-negligible when the energy density is close to Planckian values, and the dynamics quickly reduces to classical GR after the bounce as the energy density falls below Planckian values. These features facilitate readily embedding any inflationary model in an LQC background spacetime, which on the one hand allows inflationary models to become past-complete through the quantum bounce, and on the other hand also allows to study the quantum geometry signatures coming from LQC which leave distinct signatures in the CMB that have been systematically explored in the past decade \cite{LQC_CMB}.
We start with the effective Hamiltonian of LQC in the $\bar \mu$ scheme for an FRW spacetime coupled to a scalar field, which is given by
\begin{equation}
    \label{Ham_polymerized}
\mathcal{H_\mathrm{eff}}=-\frac{3v}{8\pi G \gamma^2 \lambda_b^2} \sin^2(\lambda_b b) + \frac{p_\phi^2}{2v} + v\,V(\phi) \approx 0,
\end{equation}
where the phase space consists of four canonical variables $\{v,b,\phi,p_\phi\}$ satisfying the standard Poisson bracket $\lbrace b,v \rbrace = 4\pi G \gamma$ and $\lbrace \phi,p_\phi \rbrace =1$. The geometrical sector is formulated in terms of the volume $v$ and its conjugate $b$, which in the classical theory is given by the Hubble rate as $b=\gamma H$, with $\gamma$ being the Barbero-Immirzi parameter. The polymerization parameter $\lambda_b$ is related to the minimum area gap in LQG: $\lambda^2 = 4\sqrt{3}\pi \gamma  \mathrm{\ell}_{\mathrm{Planck}}^2$. Note that the vanishing of the effective Hamiltonian constraint leads to
\begin{equation}\label{critical_density}
    \rho = \frac{3}{8\pi G \gamma^2 \lambda_b^2} \sin^2(\lambda_b b) = \rho_c \sin^2(\lambda_b b).
\end{equation}
This explicitly shows that the energy density in LQC has a global upper bound given by $\rho_c=3/8\pi G \gamma^2 \lambda_b^2$ and the bounce in LQC occurs when the energy density reaches this critical value.

The canonical equations of motion obtained from the effective Hamiltonian, which incorporates quantum corrections from the underlying quantum geometry, lead to modifications in the Friedmann and Raychaudhuri equations, while the form of the conservation equation remains the same. The modified Friedmann equation is given by 
\begin{equation}
\label{Friedmann}
    H^2=\frac{\rho}{3}\bigg( 1-\frac{\rho}{\rho_c} \bigg).
\end{equation}
Here, $\rho_c$ is the critical energy density of LQC. From the above equation, it is clear that there can be two types of turning points (from contraction to expansion and vice-versa) in the evolution - those corresponding to $\rho=\rho_c$ which lead to bounces and those corresponding to $\rho=0$ which are typically associated with recollapse. The scalar field still satisfies the Klein-Gordon equation of the same form as above:
\begin{equation}
\label{KG}
    \ddot \phi +3H\dot \phi + V'(\phi)=0.
\end{equation}
We seek a potential such that the solution to the above equation is a scalar field evolving at a uniform rate, that is $\dot \phi=-\lambda$, which implies $\ddot \phi = 0$. To obtain such a potential, we substitute $\dot \phi=-\lambda$ and $\ddot \phi=0$ in the Klein-Gordon equation, which yields $3H(-\lambda) + V'(\phi)=0$. Using the modified Friedmann equation \eqref{Friedmann} to substitute the value of $H$ in terms of the energy density, we obtain
\begin{equation}
    (V'(\phi))^2=3 \lambda^2 \bigg(\rho - \frac{\rho^2}{\rho_c}\bigg) = 3 \lambda^2 \bigg( \frac{\lambda^2}{2} + V(\phi) - \frac{(\frac{\lambda^2}{2} + V(\phi))^2}{\rho_c} \bigg).
\end{equation}
The solution to the above equation is given by
\begin{equation}
    V(\phi)=-\frac{\lambda^2}{2} + \frac{\rho_c}{2} \bigg(1 - \frac{1}{2} \bigg( e^{i\sqrt{\frac{3\lambda^2}{\rho_c}} (\phi-c)} + e^{-i\sqrt{\frac{3\lambda^2}{\rho_c}} (\phi-c)} \bigg) \bigg).
\end{equation}
The integration constant $c$ can be absorbed in a redefinition of the field, to obtain
\begin{equation}\label{potential_lqc}
    V(\phi)=-\frac{\lambda^2}{2} + \rho_c \sin^2 \bigg(\frac{1}{2} \sqrt{\frac{3\lambda^2}{\rho_c}} \phi \bigg).
\end{equation}
It can be verified by an explicit calculation that $\dot \phi = -\lambda$ is an exact solution of the Klein-Gordon equation \eqref{KG} with potential \eqref{potential_lqc}.\\

\textbf{Remark 1: } We note that the potential \eqref{potential_lqc} appears to be a polymerized version of the corresponding potential \eqref{7} for uniform rate inflation in the case of GR. In particular, this is similar to the potential obtained from a field variable-based polymer quantization of the scalar field \cite{polymerfield} (in contrast to the field momentum-based polymerization, e.g. \cite{polymerfield2}), which is also similar to the hybrid natural inflation (HNI) potential \cite{HNI1}. This establishes an important connection between uniform rate inflation and polymerized scalar field in an LQC background spacetime. This indicates that a field variable-based polymer-quantized scalar field in an LQC background admits a solution in which the scalar field rolls at a uniform rate. This is also expected to hold for the HNI potential in LQC spacetime.\\

\begin{figure}[h!]
\centering
\begin{subfigure}
    \centering
    \includegraphics[width=0.47\linewidth]
    {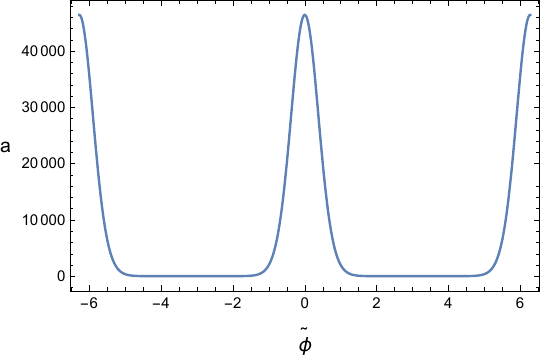}
    %\label{}
\end{subfigure}
\begin{subfigure}
    \centering
    \includegraphics[width=0.47\linewidth]
    {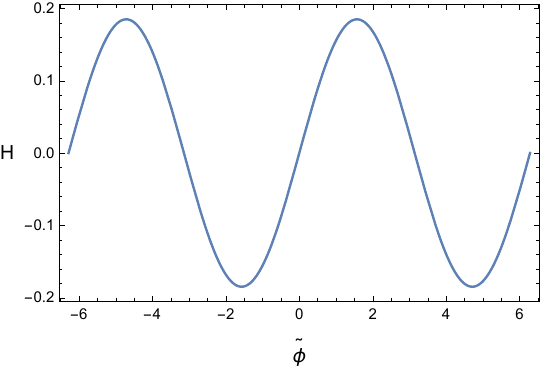}
    %\label{}
\end{subfigure}
\begin{subfigure}
    \centering    \includegraphics[width=0.47\linewidth]
    {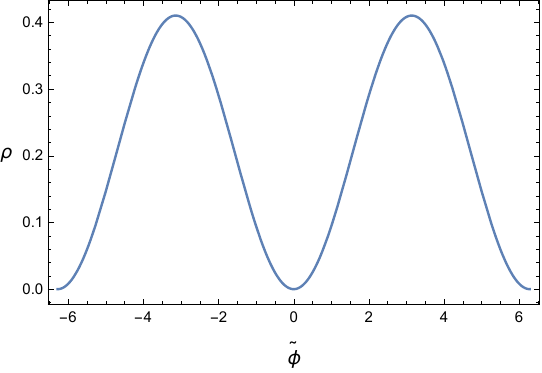}
    %\label{}
\end{subfigure}
\begin{subfigure}
    \centering
    \includegraphics[width=0.47\linewidth]
    {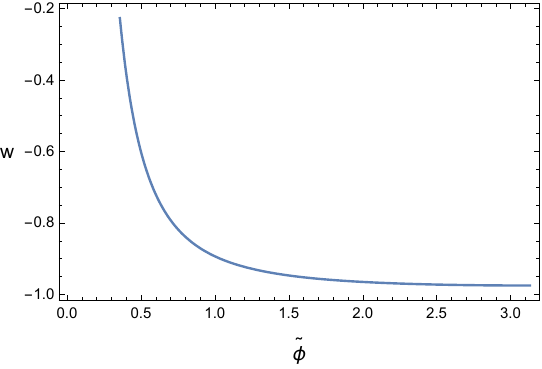}
    %\label{}
\end{subfigure}
\caption{The evolution of the scale factor $a$, the Hubble rate $H$, the energy density $\rho$ and the equation of state $w$ with respect to the rescaled scalar field $\tilde \phi  =\sqrt{\frac{3\lambda^2}{\rho_c}} \phi$ is shown, where $a_0=50$. We have chosen $\lambda=0.1$ for better viewing.}
\label{fig:evolution}
\end{figure}

\noindent
The above potential implies an energy density that is bounded
\begin{equation}\label{energy_density}
    \rho = \rho_c \sin^2 \bigg(\frac{1}{2} \sqrt{\frac{3\lambda^2}{\rho_c}} \phi \bigg).
\end{equation}
The Hubble parameter can be calculated from the modified Friedmann equation \eqref{Friedmann} as follows
\begin{eqnarray}
    H^2 &=&\frac{\rho}{3}\bigg( 1-\frac{\rho}{\rho_c} \bigg) = \frac{\rho_c}{3} \sin^2 \bigg(\frac{1}{2} \sqrt{\frac{3\lambda^2}{\rho_c}} \phi \bigg) \bigg[1-\sin^2 \bigg(\frac{1}{2} \sqrt{\frac{3\lambda^2}{\rho_c}} \phi \bigg) \bigg] \nonumber \\
    &=& \frac{\rho_c}{3} \sin^2 \bigg(\frac{1}{2} \sqrt{\frac{3\lambda^2}{\rho_c}} \phi \bigg) \cos^2 \bigg(\frac{1}{2} \sqrt{\frac{3\lambda^2}{\rho_c}} \phi \bigg) \nonumber \\
    &=& \frac{\rho_c}{12} \sin^2 \bigg(\sqrt{\frac{3\lambda^2}{\rho_c}} \phi \bigg),\nonumber \\
    H &=& \sqrt{\frac{\rho_c}{12}} \sin \bigg(\sqrt{\frac{3\lambda^2}{\rho_c}} \phi \bigg).
\end{eqnarray}
Defining $a \equiv a_0 e^\alpha$ such that $H=\dot \alpha$, we can integrate the above equation from an initial time $t_0$ to $t$ to obtain
\begin{equation}
    \alpha-\alpha_0 = \ln{\frac{a}{a_0}} = \frac{\rho_c}{6 \lambda^2} \bigg( \cos{\sqrt{\frac{3\lambda^2}{\rho_c}}\phi} - \cos{\sqrt{\frac{3\lambda^2}{\rho_c}}\phi_0} \bigg).
\end{equation}
We choose $\alpha_0=0$ and choose $\phi_0$ such that $\cos{\sqrt{\frac{3\lambda^2}{\rho_c}}\phi_0}=0$. Then the scale factor is given by
\begin{equation}
    a=a_0 e^{\alpha} = a_0 e^{\frac{\rho_c}{6\lambda^2}\cos \bigg(\sqrt{\frac{3\lambda^2}{\rho_c}} \phi \bigg)},
\end{equation}
indicating a cyclic evolution of the universe. An example of the evolution of various physical quantities with respect to the rescaled scalar field $\tilde \phi  =\sqrt{\frac{3\lambda^2}{\rho_c}} \phi$ is shown in Fig. \ref{fig:evolution} for $\lambda=0.1$. This clearly shows that the scalar field, the energy density, and the Hubble rate are bounded and oscillatory. The scale factor remains finite and non-zero, indicating the absence of any singularities. For small values of $\lambda$, the bounce appears to be a flat region, particularly in a logarithmic plot. Interestingly, the plot for the equation of state is similar to that of uniform rate inflation in general relativity with the feature that we have $w \rightarrow \infty$ when $\phi \rightarrow 0$.

\begin{figure}
    \centering    \includegraphics[width=0.89\linewidth]
    {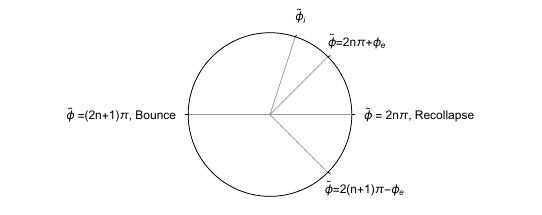}
    \caption{The cyclic evolution of the universe with respect to the rescaled scalar field  $\tilde \phi  =\sqrt{\frac{3\lambda^2}{\rho_c}} \phi$. As the scalar field is evolving backwards in our model, the evolution in the above figure is clockwise.}
    \label{fig:phi}
\end{figure}

Note that the dynamics are invariant under $\sqrt{\frac{3\lambda^2}{\rho_c}} \phi \rightarrow \sqrt{\frac{3\lambda^2}{\rho_c}} \phi+2n\pi$ where $n$ is an integer. Together with the boundedness of the physical quantities mentioned earlier, this indicates that uniform rate inflation in loop quantum cosmology results in a cyclic evolution with a compactified phase space. While the potential $V(\phi)$, the energy density $\rho$ and the scale factor are symmetric in $\phi$ (hence symmetric in time $t$), the Hubble rate is anti-symmetric in $\phi$. From the expression of the energy density and the modified Friedmann equation, it is clear that the bounce occurs when $\rho=\rho_c$, that is, when $ \sqrt{\frac{3\lambda^2}{\rho_c}} \phi = (2n+1) \pi$ where $n$ is an integer. On the other hand, the universe undergoes a recollapse at $\rho=0$ which occurs at $ \sqrt{\frac{3\lambda^2}{\rho_c}} \phi = 2n\pi$. The cyclic evolution of the universe with respect to the scalar field is illustrated in Fig. \ref{fig:phi}.\\

\noindent
\textbf{Remark 2: }We note that the present case is different from the majority of studies involving inflation in an LQC spacetime in two respects. First, we see that the bounce is dominated by potential energy (for the value of $\lambda$ consistent with observations as determined below) in contrast to the majority of studies where the bounce is found to be generally kinetic-dominated. However, this feature does not come from loop quantization, as this is also true in the corresponding GR model where the singularity is dominated by the potential energy. Second, the appearance of the polymerized potential here is in contrast to the majority of studies focusing on inflation in an LQC background where the same unpolymerized potential is often used in both GR and LQC backgrounds (see \cite{LQC_CMB} for recent works). As mentioned above, quantum gravity effects in LQC quickly become negligible after the bounce as the energy density falls below Planckian levels. This allows for readily combining LQC background spacetime with inflationary potentials used in classical GR, with only tiny signatures of quantum gravity surviving into the post-bounce inflationary phase. Most studies in LQC use the same inflationary potentials used in GR spacetimes. However, in our case, by demanding the scalar field velocity to be \textit{exactly} constant, we have required the scalar field evolution in LQC spacetime to be \textit{exactly} the same as its time evolution in classical FRW background spacetime. For this, the potential for the scalar field in LQC turns out to be the polymerized version of its GR counterpart. \\

\noindent
\textbf{Remark 3: }One may treat the dynamical system here as simply an instance of a generic matter field propagating on an effective LQC background. However, the system is special in a certain way -- we find that the polymerization of the matter sector is intimately tied to the polymerization of the background geometry, which in turn radically alters the global evolution to be cyclic. This can be seen as follows. Comparing equations \eqref{critical_density} and \eqref{energy_density} for the energy density, the polymerized terms from the matter and geometric sectors are dynamically related $\rho=\rho_c \sin^2 (\lambda_b b) = \rho_c \sin^2 (\sqrt{3\lambda^2/\rho_c}\, \phi/2)$. However, this relationship alone does not guarantee a cyclic universe (in fact, one often obtains a non-cyclic evolution with a polymerized scalar field in an effective LQC background \cite{LoiteringLQC}). The important additional factor in this particular case is that the scalar field $\phi$ evolves monotonously with a uniform negative rate $-\lambda$. Thus, the argument of $\sin (\sqrt{3\lambda^2/\rho_c}\, \phi/2)$ evolves monotonically to values $-\pi/2, -\pi, -3\pi/2, -2\pi, ....$ and so on, causing the energy density to be cyclic as per equation \eqref{energy_density}. Consequently, the universe inevitably evolves through successive bounces ($\rho=\rho_c$) and recollapses ($\rho=0$) as shown in Fig. \ref{fig:evolution} and \ref{fig:phi}. This is in contrast to standard LQC where the evolution is often not cyclic as the recollapse point $\rho=0$ is approached only asymptotically even with the inclusion of polymerized scalar fields \cite{LoiteringLQC}, unless the recollapse is explicitly induced by including special matter fields for this very purpose. In LQC, the bounce occurs when the energy density saturates to the critical energy density $\rho=\rho_c$. In our case, both the matter polymerized term $\sin(\sqrt{3\lambda^2/\rho_c}\, \phi/2)$ and the polymerized connection term $\sin^2 (\lambda_b b)$ saturate simultaneously at the bounce. Similarly, both the polymerized terms vanish simultaneously at the recollapse. In this way, the quantum effects of polymerization of geometry in LQC evolve completely in tandem with the quantum effects arising from polymer quantization of the scalar field, from the bounce to the recollapse and to the bounce again. Thus, the polymerization of the scalar field in this manuscript is special in this sense. As mentioned earlier, the quantum effects of polymerization (in this case polymerization of both the gravitational variables as well as of the potential) become negligible when the energy density falls below Planckian levels. Thus, the polymerization is expected to produce a very small effect on the CMB parameters compared to those predicted from GR.
\\

We now calculate CMB parameters. As noted above, the polymerization of matter in this case is intimately tied to the quantum geometric effects manifest in the polymerization of the geometry of the background spacetime. This is distinct from the standard LQC scenario in which the matter content simply evolves on the effective LQC spacetime. In particular, in standard LQC the matter content is often Fock-quantized which doesn't necessarily bear any specific relation to the loop quantization of the background geometry. In this manuscript, we also do not include additional scalar fields to drive post-bounce inflation. Instead, the same scalar field with the above-mentioned polymerization tied to the polymerization of the background, which governs the cyclic evolution of the background spacetime, is also responsible for the inflationary phase after the bounce. Thus, we cannot simply use the perturbation schemes developed in standard LQC for Fock-quantized matter fields for the evolution of the perturbations in this case. A careful analysis of cosmological perturbations at the effective geometry level in the present case would require the development of the underlying theory of perturbations of a polymer-quantized scalar field on a polymer-quantized discrete geometry, and then obtaining the effective equations for perturbations from that. This is outside the scope of the current work. In the absence of a rigorous framework for the evolution of cosmological perturbations suitable to the present case, we limit ourselves to estimating the CMB parameters by including only the leading-order corrections from the modified background dynamics compared to GR. Thus, our results for the CMB parameters are preliminary.

First, we consider the regime with $\ddot a>0$. The second derivative of the scale factor is given by
\begin{equation}
    \ddot a = a_0 e^{\alpha} (\ddot \alpha + \dot\alpha^2)= a\bigg( -\frac{\lambda^2}{2}\cos \bigg(\sqrt{\frac{3\lambda^2}{\rho_c}} \phi \bigg) + \frac{\rho_c}{12} \sin^2 \bigg(\sqrt{\frac{3\lambda^2}{\rho_c}} \phi \bigg) \bigg).
\end{equation}
To find the regime where $\ddot a>0$, we have to obtain the roots of the bracketed expression in the above equation. We define
\begin{equation}
   \tilde{\phi}=\sqrt{\frac{3\lambda^2}{\rho_c}} \phi,
\end{equation}
and use the shorthand notation $\cos{\tilde{\phi}}= \mathcal{C}$, $A=\lambda^2 /2$ and $B=\rho_c /12$. Substituting $\sin^2{\tilde{\phi}}=1-\cos^2{\tilde{\phi}}$, the condition $\ddot a>0$ translates to $ B \mathcal{C}^2 +A\mathcal{C} -B<0$, which is a quadratic that opens upward. Its roots are given by $\mathcal{C}_{\pm} = (-A \pm \sqrt{A^2 + 4B^2})/2B$. Clearly $\mathcal{C}_{-}<-1$ and $0<\mathcal{C}_{+}<1$. Since $|\mathcal{C}|=|\cos{\tilde{\phi}}| \leq 1$, the condition $\ddot a >0$ translates to $\mathcal{C} \in (-1,\mathcal{C}_{+})$. Thus, the inflationary phase must exist within the range
\begin{equation}
    \tilde{\phi} \in ( \arccos{\mathcal{C}_{+}}+2\pi n, 2\pi - \arccos{\mathcal{C}_{+}}+2\pi n),
\end{equation}
where $n$ is an integer and $\mathcal{C_{+}}$ is given by
\begin{equation}
    \mathcal{C}_{+} = \sqrt{1+\bigg(\frac{3\lambda^2}{\rho_c}\bigg)^2}-\frac{3\lambda^2}{\rho_c}.
\end{equation}

Note that the scalar field evolves backwards with a uniform rate, thus, the inflationary phase ends when $\tilde{\phi}=\tilde{\phi}_e=\arccos{\mathcal{C}_{+}}$, assuming we are in the $n=0$ branch. Thus, when $\phi>\phi_e = \sqrt{(\rho_c/3\lambda^2)}\tilde{\phi}_e $, inflation is happening with $\ddot a>0$. Assuming that reheating happened  when the scalar field attained the value $\phi=\phi_e$ for the  recovery of the hot big bang era. In order for the universe to leave inflation with sufficient e-foldings, say $\Delta\alpha = 60$, the initial scalar field at horizon exit can be calculated as follows
\begin{equation}\label{efoldings}
    60= \alpha_e-\alpha_i = \frac{\rho_c}{6\lambda^2}\bigg( \cos{\sqrt{\frac{3\lambda^2}{\rho_c}}\phi_e} - \cos{\sqrt{\frac{3\lambda^2}{\rho_c}}\phi_i} \bigg) = \frac{\rho_c}{6\lambda^2}\bigg( \mathcal{C}_{+} - \cos{\sqrt{\frac{3\lambda^2}{\rho_c}}\phi_i} \bigg).
\end{equation}
This gives
\begin{equation}\label{cosxi}
    \cos{\sqrt{\frac{3\lambda^2}{\rho_c}}\phi_i} = \sqrt{1+\bigg(\frac{3\lambda^2}{\rho_c}\bigg)^2}-\frac{363\lambda^2}{\rho_c}
\end{equation}
We now use the $\delta N$ formalism \cite{5,6,7} to calculate the primordial curvature perturbations. We have $\delta N = -\delta \alpha$, thus
\begin{equation}
    \delta N = -\frac{\partial \alpha}{\partial \phi} \delta \phi = \frac{\rho_c}{24\pi \lambda} \sin^2 \bigg(\sqrt{\frac{3\lambda^2}{\rho_c}} \phi \bigg).
\end{equation}
As mentioned above, perturbation equations for precisely calculating the fluctuation $\delta \phi$ in the polymerized scalar field on a polymerized LQC background are not available. Thus, we limit ourselves to including only the leading-order corrections from modifications to the background dynamics in LQC. Our estimates are thus preliminary in this sense. We estimate $\delta \phi \approx \frac{H}{2\pi}$ where the Hubble rate is determined by the modified Friedmann equation for the LQC background spacetime. This is justified as the lowest-order calculation that includes only the effect of background dynamics in LQC (see e.g. \cite{Mielczarek}). The spectrum $P_R$ is then given by
\begin{equation}
    P_R = (\delta N)^2 = \frac{\rho_c^2}{576\pi^2 \lambda^2} \bigg(1- \cos^2 \bigg(\sqrt{\frac{3\lambda^2}{\rho_c}} \phi \bigg) \bigg)^2 = \frac{\rho_c^2}{576\pi^2 \lambda^2} \bigg(1 - \frac{36\lambda^4}{\rho_c^2} \alpha^2 \bigg)^2
\end{equation}
Comparing with the Cosmic Microwave Background (CMB) normalization $\zeta=5 \times 10^{-5}$, we obtain
\begin{equation}\label{zeta}
    \zeta = \frac{\rho_c}{24\pi \lambda} \sin^2 \bigg(\sqrt{\frac{3\lambda^2}{\rho_c}} \phi_i \bigg) = \frac{\rho_c}{24\pi \lambda} (1- \cos^2 {\tilde{\phi}_i}) 
\end{equation}
Substituting the value of $\cos{\tilde{\phi}_i}$ from equation \eqref{cosxi}, this yields
\begin{equation}
    \zeta= \frac{\lambda}{4\pi} \bigg(121\sqrt{1+\bigg(\frac{3\lambda^2}{\rho_c}\bigg)^2}-\frac{21963\lambda^2}{\rho_c} \bigg).
\end{equation}
Solving this equation, we obtain two positive values for $\lambda$: $5.19 \times 10^{-6}$ and $4.75 \times 10^{-2}$.

We now calculate the spectral index $n_s$, which is given by
\begin{equation}
    n_s=1+\frac{d \ln{P_R}}{d\ln{k}}=1+\frac{d \ln{P_R}}{d\alpha} = 1-\frac{\frac{144\lambda^4}{\rho_c^2}\alpha_i}{1-\frac{36\lambda^4}{\rho_c^2} \alpha_i^2}
\end{equation}
Using equation \eqref{zeta}, we can substitute $1- \cos^2 {\tilde{\phi}_i} = (12\pi \lambda /\rho_c) \times 10^{-4}$, which yields the spectral index
\begin{equation}
    n_s=1- \frac{20000\lambda}{\pi}\bigg( \sqrt{1+\bigg(\frac{3\lambda^2}{\rho_c}\bigg)^2}-\frac{363\lambda^2}{\rho_c} \bigg).
\end{equation}
For $\lambda = 5.19 \times 10^{-6}$ we obtain $n_s = 0.966942$, which is consistent with the observations, while for the other value of $\lambda$ the $n_s$ is too large. Thus we fix the parameter $\lambda$ to be
\begin{equation}
    \lambda = 5.19 \times 10^{-6}.
\end{equation}
Next, we examine the tensor-to-scalar ratio. As in the case of the scalar power spectrum, due to the lack of evolution equations for perturbations suitable to the present case, we calculate the tensor power spectrum only including the corrections to background dynamics from due to the modified Friedmann equation in LQC. The spectrum for tensor perturbations including only the background dynamics corrections is given by \cite{Mielczarek}:
\begin{equation}
    P_T=8\bigg( \frac{H}{2\pi} \bigg)^2.
\end{equation}
The tensor-to-scalar ratio comes out to be
\begin{equation}
    r=\frac{P_T}{P_R} = \frac{96\lambda^2}{\rho_c} \frac{1}{1-\cos^2{\tilde{\phi}_i}},
\end{equation}
which turns out to be $0.13$ for the value of $\lambda$ selected by the spectral index. This does not satisfy the observational constraints on $r$. 

\begin{figure}
    \centering    \includegraphics[width=0.69\linewidth]
    {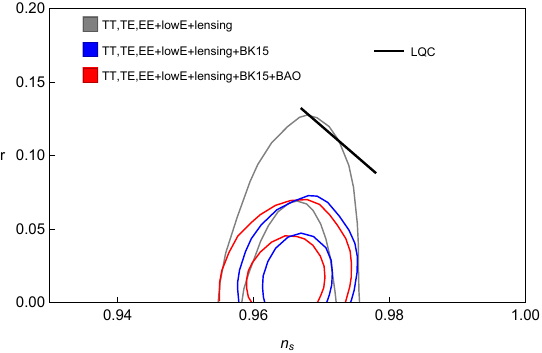}
    \caption{The tensor-to-scalar ratio and the spectral index. The constraints are from Planck data \cite{Planck}. The solid black line represents the values of tensor-to-scalar ratio and spectral index that are possible to obtain from LQC.}
    \label{fig:rVSns}
\end{figure}

However, we could potentially satisfy the constraints if we assume that inflation ends earlier than $\phi=\phi_e$, say at $\phi=\phi_f$ where $\phi_f>\phi_e$, such that $\phi_i$ is also shifted to an earlier time, thus increasing the value of the factor $(1-\cos^2{\tilde{\phi}_i})$. However, the spectral index $n_s$ must also remain within the constraints. If the e-foldings between the end of infaltion and $\phi=\phi_e$ are given by $\Delta\alpha_{fe} = \alpha_e - \alpha_f = m$ (say), then we have $\alpha_e - \alpha_i = 60+m$ and equation \eqref{efoldings} gets modified, leading to a modified value for $\cos{\tilde{\phi}_i}$ given by
\begin{equation}\label{cosxi_modified}
    \cos{\sqrt{\frac{3\lambda^2}{\rho_c}}\phi_i} = \sqrt{1+\bigg(\frac{3\lambda^2}{\rho_c}\bigg)^2}-\frac{363\lambda^2}{\rho_c}-\frac{6 m \lambda^2}{\rho_c}.
\end{equation}
This changes both the spectral index and the tensor-to-salar ratio. We plot the spectral index and the tensor-to-scalar ratio for different values of $m$ ranging from $0$ to $30$ in Fig. \ref{fig:rVSns}, showing that it is barely possible to satisfy the constraints. We note that the results for the spectral index and the tensor-to-scalar ratio are strikingly similar to those obtained for uniform rate inflation in GR obtained in \cite{5}. We again emphasize that our results for the CMB parameters in this case are preliminary as they include only the corrections to the background dynamics. At the level of this preliminary analysis, we found a broad agreement with the observations. However, a thorough analysis of cosmological perturbations is expected to reveal tiny departures from the standard model in the angular power spectrum and the magnitude of non-Gaussianities. Further, the inflationary imprints of the cyclicity of the universe over multiple bounce cycles also need to be studied. However, to accurately include the corrections to the perturbations in this particular case would require the study of polymer perturbations on the loop-quantized discrete geometry of LQC, lacking which it is only possible to include the corrections to the background dynamics at present.

The differences from and similarities with the case of uniform rate inflation in GR can be understood as follows. As noted above, the background dynamics is radically modified compared to GR, leading to a cyclic universe in this case as shown in Fig. \ref{fig:evolution}. This is due to two underlying reasons. First, the strong quantum gravity effects arise from underlying discrete quantum geometry when the energy density becomes Planckian, leading to a quantum bounce which resolves the classical singularity. Second, owing to the modifications introduced in the Friedmann equations by quantum geometry, the scalar field potential for obtaining uniform rate inflation in loop quantum cosmology turns out to be very different from that required in GR. In fact, as noted above, the potential \eqref{potential_lqc} appears to be a polymerized version of the corresponding potential \eqref{7} in GR. Thus the background dynamics is radically different from its classical counterpart. As noted in the third remark above, this cyclic evolution is due to the specific polymerization of the potential obtained in this study, which in turn results from demanding the scalar field in LQC spacetime to \textit{exactly} mimic its evolution in classical spacetime. However, an important aspect of loop quantization is that the effects of loop quantization are often unnoticeable in regimes where the energy density and curvature are far lower than Planckian values and the universe is macroscopic. In these regimes, LQC trajectory very closely approximates the classical evolution. The effects of quantization become apparent only when one looks at the global evolution which gets radically modified due to the presence of the quantum bounce. This explains the results here. While the global evolution of the background is radically modified, there is only a small effect on the CMB parameters, which can be uncovered with further detailed investigation.

\section{Summary}

We investigated uniform rate inflation in loop quantum cosmology with a scalar field. Unlike the majority of studies focusing on inflation in an LQC spacetime, the bounce in this case is potential energy dominated. Moreover, unlike previous studies, most of which embed the same inflatonary potentials in LQC as used in GR, which typically only slightly alters the inflationary phase as the quantum corrections coming from the bounce become negligible in the low curvature inflationary phase, here we have demanded the scalar field evolution in LQC spacetime to exactly mimic its evolution in the GR spacetime, i.e. an \textit{exactly} uniform rate inflation. We found that the potential for an \textit{exactly} uniform rate inflation in loop quantum cosmology closely resembles the potentials for a polymer quantized scalar field and that of hybrid natural inflation (HNI). This potential is the polymerized version of the corresponding potential for uniform rate inflation in a classical GR background spacetime. Matter polymerized on the lines of gravitational variables has rarely been considered in LQC, except in a few works where it has been put in deliberately. This is the first time that the polymerization of matter quantities has resulted purely from phenomenological considerations in LQC without putting it in by hand. Moreover, these phenomenological considerations select a specific polymerization of the scalar field which is intimately tied to the quantum geometry effects in the background spacetime in LQC. In particular, the magnitude of corrections due to polymerization of the scalar field evolves in tandem with the magnitude of quantum corrections to the background geometry reflected in the polymerization of the connection and the modified Friedmann equation. As a result of this and the particular dynamics of the scalar field (i.e. its uniform evolution), the universe turns out to be cyclic with identical cycles of expansion and contraction joined through the quantum bounce which replaces the classical singularity. This is presented in Fig. \ref{fig:evolution} and Fig. \ref{fig:phi}. Thus, we find that seemingly unrelated concepts such as HNI inflation, polymer quantization of the scalar field, uniform rate inflation and cyclic evolution turn out to be closely related when the background spacetime is loop quantized. A preliminary analysis of the CMB parameters is also presented by including only the corrections from the modified background dynamics due to LQC, as the equations for the evolution of perturbations of a polymer scalar field in a loop quantized background have not yet been developed. The range of permitted values of the field for sufficient inflation is obtained. The spectral index and the tensor-to-scalar ratio permitted by uniform rate inflation in LQC are estimated and compared with observations in Fig. \ref{fig:rVSns}, showing that the model is viable and permits a realistic universe that emerged from an early inflationary era. As expected, the predictions are very close to those obtained for uniform rate inflation in a GR background spacetime, since the quantum gravity modifications become negligible in the low curvature inflationary phase, leaving only a tiny imprint. While the inflationary phase is only mildly affected, combining uniform rate inflation with loop quantum cosmology has profound impact on the global evolution of the spacetime, which becomes cyclic. At this stage, the estimates of CMB parameters are preliminary because of the lack of a suitable framework for the evolution of polymer matter perturbations on quantum geometry. We leave a full analysis of the perturbations to future studies. It is interesting to note that a single scalar field can simultaneously govern the cyclic evolution of the universe as well as drive the post-bounce inflationary phase. Moreover, the present manuscript highlights the linkages between various seemingly unrelated paradigms, some of which may turn out to be deep. The model is simple, yet it unites several interesting concepts such as the cyclic universe, uniform rate inflation, polymer quantized scalar field and hybrid natural inflation. It would be interesting to further investigate these connections in future studies.

\section*{Acknowledgments}
 
BCP is thankful to CERN Switzerland for supporting Visiting Scientist program. SS is supported by Guru Jambheshwar University of Science \& Technology, India with Seed Money Grant vide Endst. No. DR\&D/2025/103-10.

\end{document}